\documentclass[
reprint,
superscriptaddress,
 amsmath,amssymb,
 aps,
]{revtex4-2}

\usepackage{xcolor}
\usepackage{graphicx}
\usepackage{dcolumn}
\usepackage{bm}
\usepackage{soul}
\usepackage{hyperref}
\hypersetup{colorlinks=true,linkcolor=blue,citecolor=blue}

\begin{document}


\title{In Situ Observation of Proton-Induced DNA fragmentation in the Bragg Peak: Evidence for Protective Role of Water.
}

\author{R. Liénard}%
\affiliation{Univ. Bordeaux, CNRS, LP2I, UMR 5797, F-33170 Gradignan, France}%

\author{P. Barberet}
\affiliation{Univ. Bordeaux, CNRS, LP2I, UMR 5797, F-33170 Gradignan, France}%

\author{K. Chatzipapas}%
\affiliation{Delft Univ. of Technology, Radiation Science and Technology, Faculty of Applied Sciences, 2629 JB Delft, Netherlands}%

\author{G. Devès}

\affiliation{Univ. Bordeaux, CNRS, LP2I, UMR 5797, F-33170 Gradignan, France}%

\author{T. Dhôte}%
\affiliation{Univ. Bordeaux, CNRS, LP2I, UMR 5797, F-33170 Gradignan, France}%

\author{T. Guérin}%
\affiliation{Univ. Bordeaux, CNRS, LOMA, UMR 5798, F-33400 Talence, France}%

\author{H. Seznec}%
\affiliation{Univ. Bordeaux, CNRS, LP2I, UMR 5797, F-33170 Gradignan, France}%

\author{F. Gobet}
 \email{gobet@lp2ib.in2p3.fr}

\affiliation{Univ. Bordeaux, CNRS, LP2I, UMR 5797, F-33170 Gradignan, France}%

\date{\today}

\begin{abstract}
We report {\it in situ } single-molecule measurements of proton-induced double-strand breaks (DSBs) in DNA immersed in water, using real-time fluorescence tracking along the entire proton path, including the Bragg peak region. By chemically suppressing radical-mediated processes, we isolate direct DNA damage mechanisms and determine DSB cross-sections as a function of depth. Near the Bragg peak, we observe a tenfold reduction in DSB cross-sections in aqueous DNA compared to dry DNA, providing quantitative evidence for the protective role of water. These findings highlight the importance of intermolecular energy dissipation in mitigating radiation-induced damage in condensed biological matter, with implications for radiobiology and proton therapy modeling.

\end{abstract}

\maketitle

{\it Introduction.--} Quantifying the effects of ionizing radiation  on DNA in aqueous environments is crucial for unraveling the mechanisms of radiation-induced biological damage 
\cite{chicana2023radical, sanche2009beyond, gopakumar2023radiation, shepard2023electronic}, with profound implications for radiotherapy 
\cite{durante2021physics,vozenin2022towards} and radiation protection \cite{durante2008heavy}. 
Among ionizing particles, protons and heavy ions are particularly relevant due to their distinctive energy deposition profile in matter, which features a sharp maximum-known as the Bragg peak- just before they are stopped \cite{datz2013condensed,shepard2023electronic, durante2010charged}. This localized energy release enables precise  tissue targeting in applications like proton therapy. 
Such particles can induce various type of genetic damage, particularly double strand breaks (DSBs), 
which disrupt both DNA strands and lead to harmful mutations 
\cite{asaithamby2011unrepaired,vitti2019radiobiological}. 

Proton-induced DNA damage has been studied in liquid water outside the Bragg peak \cite{siefermann2010binding,alizadeh2012precursors,vyvsin2015proton,ohsawa2022dna}. In these conditions, DSBs arise from either indirect chemical effects of radicals generated by water radiolysis, 
or direct energy transfer to DNA. 
Indirect processes increase DNA damage yields in aqueous environments and can be modulated using radical scavengers \cite{vyvsin2015proton}. In contrast, water's role in direct damage pathways under condensed phase conditions remains poorly understood.

Recent developments in cluster and molecular physics have highlighted the crucial role of water
in mediating energy transfer between biomolecules and their environment through intermolecular interactions, 
such as hydrogen bonding\cite{liu2006collision,markush2016role,domaracka2012ion,barc2014multi,johny2024water,abdoul2015velocity,berthias2015proton,kocisek2016microhydration,richter2018competition,jahnke2010ultrafast,ren2018experimental,wang2020water,milosavljevic2020oxygen}.
Experiments on small biomolecules with limited hydration show
significantly reduced radiation damage \cite{liu2006collision,markush2016role,domaracka2012ion,barc2014multi,johny2024water} - suggesting a protective role of water. 
Conversely, energy released from excited or ionized water molecules can also be transferred to nearby biomolecules, 
suggesting a possible catalytic role of water  in damage induction \cite{jahnke2010ultrafast,ren2018experimental,wang2020water,milosavljevic2020oxygen}. However, these experiments typically involve small systems and only a few water molecules, making their relevance to full DNA in the condensed phase uncertain.

Whether water predominantly protects or enhances direct DNA damage at this scale
remains an open question. To address it, we need to measure and compare DSB cross-sections in dry and 
aqueous DNA, in the presence of scavengers to minimize the radical effects, particularly in the Bragg peak region \cite{shepard2023electronic}, 
where diverse ionization and excitation pathways offer optimal conditions for observing intermolecular interaction effects. 
While  DSB cross-sections in dry DNA by protons in vacuum have been reported\cite{souici2017single}, 
experimental limitation have so far precluded equivalent measurements in liquids within the Bragg peak, mainly due to the difficulty of extracting biological 
material from the ultrathin liquid layer required for standard {\it ex situ} analysis \cite{vyvsin2015proton,frame2022proton}. 
Furthermore, to date, no single-molecule {\it in situ} observation of DNA double-strand breaks enabling exploration of the Bragg peak region has been reported.

In this work, we overcome these limitations 
using an
 {\it in situ} single-particle tracking approach to measure DSBs induced by proton 
along their full trajectory in water- from entry into the water target to their stopping in the Bragg peak region. 
Combining nuclear and photonic techniques with soft matter analysis, we isolate the direct contribution to DNA fragmentation by minimizing the radical-mediated effects. We then quantify DSB cross-sections 
as function of depth. Our experimental results reveal a substantial reduction in DSB cross-sections at the Bragg peak in aqueous DNA compared to dry DNA, 
demonstrating that, at the scale of condensed matter, water's protective effect outweighs any catalytic role in direct DNA damage.

\begin{figure}[h]
\includegraphics[width=8.6cm,trim=0 50 0 0, clip=true]{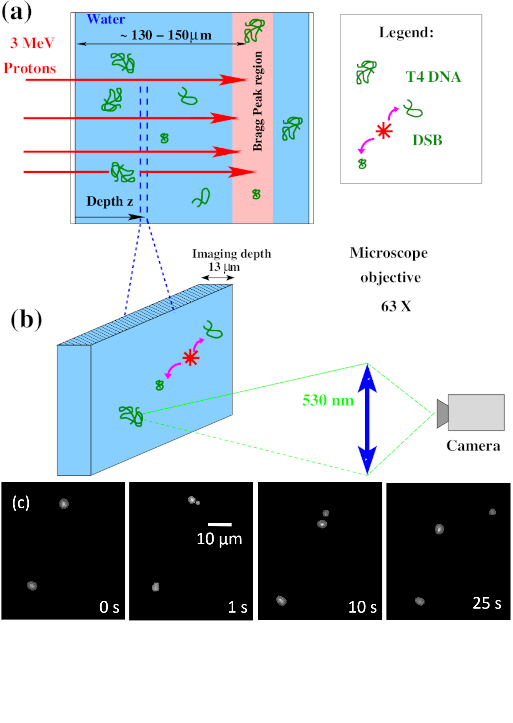}
\caption{\label{Fig1} Principle of the {\it in situ} observation of DSB DNA fragmentation (a) A DNA solution is irradiated with 3 MeV protons up to the Bragg peak region. (b) DNA dynamics at depth z are captured via fluorescence microscopy within a 13 µm imaging depth. (c)  Sequential snapshots of DNA after a 400 ms irradiation of 6,000 protons µm$^{-2}$ applied at time t = 0 s. Two T4 DNA molecules are visible, with the upper one undergoing a DSB event resulting in a multiplicity of 2.}
\end{figure}

{\it In situ method.--} The experiments are carried out on a beamline of the AIFIRA facility at the Laboratoire de Physique des 2 Infinis in Bordeaux \cite{muggiolu2017single}. A 3 MeV proton beam is focused on the target at normal incidence, as illustrated in Fig. 1(a). The beam is carefully controlled to deliver a fluence of up to 6,000 protons µm$^{-2}$ in a uniform field of $300\times250$ µm$^2$, with irradiation lasting less than 400 ms. 

The target is a film of aqueous buffer solution confined between two 4 µm thick polypropylene sheets, positioned in air 1 mm behind a 200 nm-thick Si$_3$N$_4$ window. The target is about 200 µm thick, which includes the Bragg peak at a depth of 142 µm for 3 MeV protons \cite{pstar}. The buffer solution (pH = 8.4) contains 45 mM Tris-Boric acid to keep DNA in a soluble state, as well as 1 mM EDTA to prevent enzymatic degradation of DNA \cite{ogden19878}. Additionally, 140 mM $\beta$-Mercaptoethanol is added to scavenge OH$^.$ radicals \cite{stanton1993protection}, which significantly reduces the chemical contribution of free radicals to DNA fragmentation \cite{vyvsin2015proton,stanton1993protection}. 

The DNA in the solution consists of 169,000 base pairs (bp) of linear double-stranded DNA from phage T4 \cite{miller2003bacteriophage}, labeled with YOYO-1 fluorescent dye \cite{larsson1995characterization}. Labeling is achieved by incubating the DNA with the dye at a ratio of one dye molecule per four base pairs. DNA dynamics are tracked  using an inverted fluorescence microscope. A 470 nm diode light is reflected by a dichroic mirror into the microscope, where a 63$\times$ objective lens illuminates the sample and collects fluorescence emitted at 530 nm (Fig. 1(b)). The emitted signal is recorded by a camera, providing real-time imaging of DNA strands at a depth z in the target. The imaging depth, approximately 13 µm, allows for macromolecule tracking despite their longitudinal Brownian motion. Additional information on the experiment are detailed in Sec. I of the Supplemental Material~\cite{supmat}\nocite{crocker1996methods,doi1988theory,redner2001guide,rotne1969variational,wajnryb2013generalization,engl1996regularization,buccini2017iterated,incerti2010geant4,incerti2018geant4,nikjoo2016radiation,friedland2003simulation,chatzipapas2024development}.

{\it DNA solution and event characterization.--} To ensure accurate analysis, the DNA sample is diluted so that no more than 5 to 6 macromolecules are present within a $127\times127$ µm$^2$ field of view. The T4 DNA diffusivity is measured by tracking its Brownian motion, yielding a value of $D$ = 0.32 $\pm$ 0.02 µm$^2$s$^{-1}$, which corresponds to a radius of gyration R$_G$ = 1 µm based on the Zimm model \cite{smith1996dynamical}.

T4 DNA is linear, so when a DSB occurs, the resulting fragments move apart due to Brownian motion. For the slowest symmetric fragmentation, where the DNA breaks into two pieces, the diffusivity of the fragments is around 0.50 µm$^2$s$^{-1}$, which means that it takes roughly 6 to 7 s for the fragments to separate by more than five times the radius gyration, as seen in the video plane (see Sec. II of the Supplemental Material and three illustrative videos \cite{supmat}). A representative example of this process is shown in Fig. 1(c), where at t=0, a proton pulse is applied. One DNA molecule undergoes fragmentation into two pieces, indicating a DSB. In total, 8681 T4 DNA molecules (referred to as events) were tracked, with 1312 of them fragmenting into at least two pieces.

 After processing the images with a machine-learning algorithm designed to identify DNA in each frame \cite{berg2019ilastik}, the DNA fragment dynamics are extracted event-by-event using a home made analysis program. Key recorded parameters include the number of fragments (multiplicity) for each DNA molecule, their Brownian trajectories within the imaging plane, their fluorescence intensity, and the average depth $z$ at which the video is acquired within the liquid layer. The shortest detectable fragment size, L$_{short}$ plays a crucial role in the analysis. Fragments with luminosity below the background noise are undetected, introducing a bias that requires correction. Since DNA brightness is proportional to the number of fluorescent dyes bound to the macromolecule -and therefore to the number of base pairs-, L$_{short}$ is determined by comparing the DNA fluorescence intensity to the local background in each image.
 Variation in background fluorescence and DNA intensity due to differences in imaging depth result in different L$_{short}$ values for each DNA molecule.
 The length distribution P(L$_{short}$), constructed from the entire dataset, reveals that in 90\% of events, L$_{short}$ ranges from 5 kbp to 33 kbp, with an average value of around 20 kbp. This suggests that, on average, about 75\%  of the T4 DNA is effectively probed through fragmentation events resulting from a single DSB; see Sec. II.D of the Supplemental Material for further details  \cite{supmat}.

{\it Reduced DNA direct damage yield in water.--} To probe how DNA damage varies with proton penetration depth, we analyze datasets acquired at different imaging depths along the z-axis of the water target, covering both pre-Bragg and Bragg peak regions. We begin by analyzing a dataset in which DNA molecules are irradiated at depths between 25 and 90 µm, under varying proton fluences. This region, well upstream of the Bragg peak, corresponds to 3 MeV protons that slow down by 1.5 MeV, with an energy loss varying slightly between 12 and 17 keV/µm \cite{pstar}. Over this range,  the DSB cross-section remains approximately constant. This allows us to perform fluence-dependent measurements, each based on several hundred DNA molecules to ensure statistical significance. Figure 2 shows the distribution of multiplicities as a function of proton fluence. As expected, the probability of DNA fragmentation (multiplicity m $\ge$ 2) increases with fluence. At the highest fluence of 6,000 protons µm$^{-2}$, 35$\%$ of T4 DNA molecules break into at least two pieces.

\begin{figure}[h]

\includegraphics[width=8.6cm,trim=0 0 0 0, clip=true]{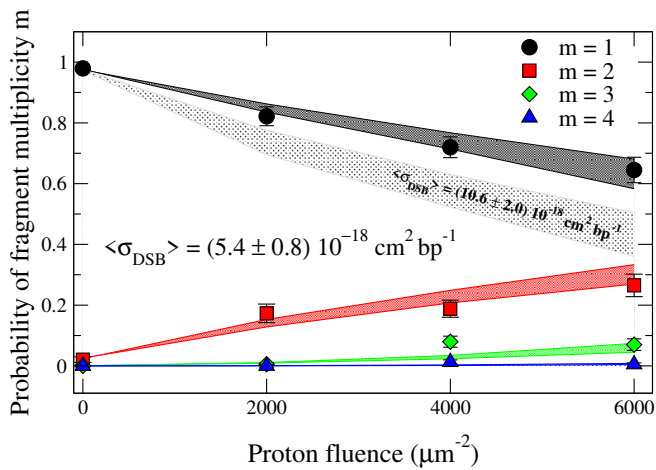}

\caption{\label{Fig2} Multiplicity distributions for 1.5-3 MeV proton - T4 DNA collisions in water target within the 25-90 µm depth range. Experimental results are represented by symbols. The high-density shaded areas correspond to fits using the fragmentation model with $\langle\sigma_{DSB}\rangle$=(5.4 $\pm$ 0.8)$\times$10$^{-18}$cm$^2$bp$^{-1}$. The low density shaded area provides prediction for m=1 (no fragmentation) with $\langle\sigma_{DSB}\rangle$=(10.6 $\pm$ 2.0)$\times$10$^{-18}$cm$^2$bp$^{-1}$ as expected for dry DNA.}

\end{figure}

We introduce a theoretical analysis to extract the DNA double-strand break (DSB) cross-section from measured fragment multiplicity distributions, explicitly incorporating experimental detection thresholds and aiming to reduce associated biases. DNA is modeled as a linear chain of L base pairs linked by phosphate-sugar bonds, each with a break probability $p=\sigma_{bond}\times\Phi$, where $\Phi$ is the proton fluence and $\sigma_{bond}$ the bond-breaking cross-section.
The simulation proceeds in three steps, (illustrated in Fig. 3 for 4 events): (1) random assignment of bond breaks along the DNA; (2) application of a size detection threshold for each event, sampled from the experimentally length distribution P(L$_{short}$), to exclude undetectable fragments; (3) calculation of the detected fragment multiplicity.
For each parameter set, 10,000 events are simulated. The DSB cross-section per base pair, $\langle\sigma_{DSB}\rangle$, is obtained by adjusting $\sigma_{bond}$ to minimize the residual between simulated and experimental multiplicity distributions. This procedure, which incorporates detection limits at each step, enables robust determination of $\langle\sigma_{DSB}\rangle$.

\begin{figure}[h]

\includegraphics[width=8.6cm,trim=0 0 0 0, clip=true]{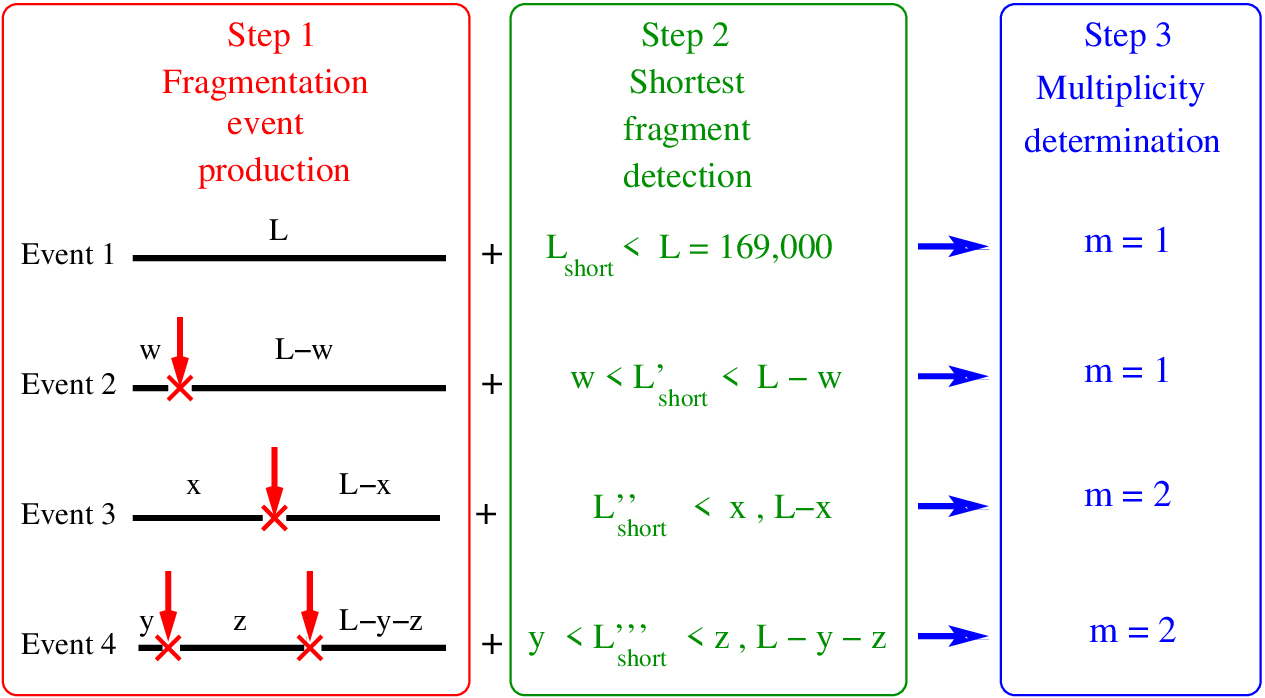}

\caption{\label{Fig2b} Schematic representation of the fragmentation analysis applied to four simulated events, showing: (1) random bond breaking, (2) application of size detection thresholds, and (3) resulting detected fragment multiplicities. The letters w, x, y, and z denote four distinct fragment sizes.}
\end{figure}

As shown in Fig.2, excellent agreement is achieved with $\langle\sigma_{DSB}\rangle = (5.4 \pm 0.8)\times10^{-18}$ cm$^2$bp$^{-1}$. This cross-section is 50\% lower than the value of $(10.6 \pm 2.0)\times10^{-18}$ cm$^2$bp$^{-1}$  estimated by Souici et al for dry DNA in vacuum at proton energies between 1.5 and 3 MeV, corresponding to the energy range of this study \cite{souici2017single}. Using Souici et al.'s cross-sections in our experiment would predict a fragmentation yield of 50-60\% at a fluence of 6,000 protons µm$^{-2}$, which is inconsistent with the experimental data, as shown in Fig. 2. This discrepancy indicates that the cross-section for direct DNA fragmentation in liquid is smaller than for dry DNA.

To assess the reliability of our result,
we examined several factors that could influence the measured cross-sections. 
First, we verified that fluorescent dyes did not affect the probability of DNA fragmentation. 
Additional experiments were performed at lower dye loadings, using YOYO-1 dye at ratios as low as 1 dye per 12 base pairs, at which point the DNA fluorescence became insufficient for quantitative analysis. In this range, YOYO-1 molecules bind to DNA in different modes, shifting from mono-intercalation (1 dye per 4 base pairs) to a bis-intercalation (1 dye per 12 base pairs )\cite{larsson1995characterization,kucharska2021two}. Despite this changes in binding modes, no significant variation in the DSB cross-section was observed.  
Second, we tested the effect of reducing the $\beta$-Mercaptoethanol concentration by a factor of ten, which increases the lifetimes of free radical by the same order of magnitude. This modification did not affect the DSB cross-section, confirming the negligible influence of indirect processes. Sec. III of the Supplemental Material details measured cross-sections as a function of dye charge and $\beta$-mercaptoethanol concentration  \cite{supmat}.
Third, we ensured that the fragmentation model did not introduce any bias in the cross-section extraction process. When using the DSB cross-sections published in Ref.\cite{souici2017single}, the model accurately reproduced the reported DSB production yields. 

Finally, we validated the cross-section for dry DNA at a proton energy of 3 MeV by repeating a DSB cross-section measurement following the protocol outlined by Souici et al., but with different initial plasmid topology conditions. The resulting cross-section, $\sigma_{DSB}=(10.4 \pm 1.3)\times10^{-18}$ cm$^2$bp$^{-1}$ \cite{beaudier2024quantitative}, is in close agreement with the previously reported value of $(8.9 \pm 2.2)\times10^{-18}$ cm$^2$bp$^{-1}$ \cite{souici2017single}. These cross-sections, corresponding to the proton energy at the target entrance, are already significantly higher than those measured in aqueous solution. 
Based on these checks, we conclude that the lower DSB cross-section measured in water likely reflects a reduced efficiency of direct DNA damage processes in aqueous environments compared to dry conditions.

{\it Enhanced water effect in the Bragg-Peak region.--} To further explore this scenario, we analyze a dataset capturing DNA dynamics at various depths within the 200 $\mu$m-thick target to examine the DSB cross-section specifically in the Bragg peak region. Here, a more complex interplay of ionization, excitation and relaxation mechanisms is expected, potentially altering the balance between the protective and catalytic effects of water.  Event-by-event analysis enables classification of events into distinct sub-groups based on the water depth z at which the videos are recorded. Each sub-group ($z_m$,$z_M$) includes all events for which the central imaging planes lie between $z_m$ and $z_M$, with $z_M$-$z_m$=20 $\mu$m. These depth intervals are chosen to ensure a minimum of 200 events per subgroup, thereby providing statistically robust data for extracting the raw cross-sections, $\overline\sigma_{DSB}$ using the fragmentation analysis. Due to beam time constraints, a single fluence of 2,000 protons $\mu$m$^{-2}$ was used, chosen to optimize the event statistics while minimizing total irradiation time.

The raw cross-sections shown in Fig. 4 reveal two distinct regions: a 90 µm-wide plateau where the cross-sections fluctuate around $\langle\sigma_{DSB}\rangle$, followed by a broad peak centered at approximately 125 $\mu$m. This profile closely mirrors the typical Bragg peak behavior, with a maximum cross-section of (15 $\pm$ 2)$\times$10$^{-18}$ cm$^2$bp$^{-1}$, which is 25 times lower than the (380 $\pm$ 95)$\times$10$^{-18}$ cm$^2$bp$^{-1}$ cross-section measured for dry DNA at the Bragg peak energy in Ref. \cite{souici2017single}. This marked difference provides strong evidence for a protective effect of water, although the extent of this protection in the Bragg peak region warrants further consideration. 

\begin{figure}[t]

\includegraphics[width=8.6cm,trim=0 0 0 0, clip=true]{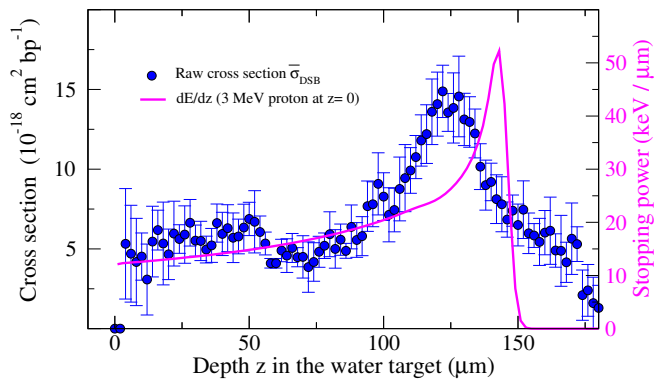}

\caption{\label{Fig3} Depth profile of the raw DSB cross-section over 20 µm depth intervals (blue circle) alongside the stopping power profile of 3 MeV incident protons (magenta line).}
\end{figure}

The data averaging over 20 µm depth intervals, combined with uncertainties in precisely locating the T4 DNA molecules due to the microscope's imaging depth, introduces low-pass filtering effects, that must be corrected.  These effects are highlighted by comparing the raw cross sections with the theoretical spatial profile of the stopping power of incident 3 MeV protons in water \cite{ziegler2010srim}, which yields a peak width of approximately 40 µm, broader than the expected 10-15 µm. Furthermore, the peak center is shifted about 20 µm closer to the entrance face than expected.

\begin{figure}[b]

\includegraphics[width=7.5cm,trim=0 0 0 0, clip=true]{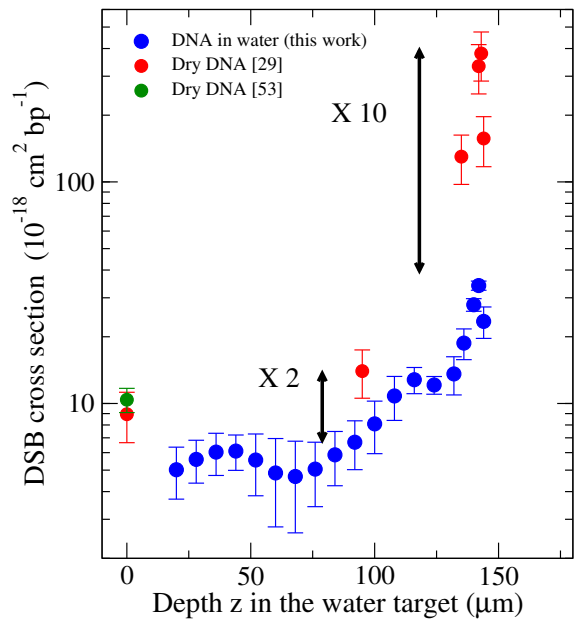}

\caption{\label{Fig4} Depth profile of the unfolded DSB cross-sections for DNA in the water target (blue circle). DSB cross-sections for dry (isolated) DNA are shown as red \cite{souici2017single} and green \cite{beaudier2024quantitative} circles. Proton energies used to measure these larger cross-sections are converted into depths within the water \cite{ziegler2010srim}, considering a proton energy of 3 MeV at the entrance face of the target.}
\end{figure}

To correct for these low-pass filtering effects, we first measure the probability per unit depth of detecting a T4 DNA molecule at its actual depth $z_0$ when the microscope images a zone centered at depth $z$. This probability is then incorporated into the reconstruction of the DSB cross-section $\sigma_{DSB}$ at a given depth using the iterative Tikhonov regularization method \cite{buccini2017iterated}. 
 
Further details can be found in Sec. IV of the Supplemental Material  \cite{supmat}. The corrected DSB cross-section profile is shown in Fig. 5 and compared with the values for dry DNA measured by Souici et al. \cite{souici2017single}, after converting proton energies into water depths using the SRIM code \cite{ziegler2010srim}. The peak better matches the expected Bragg profile and confirms that, even at maximal energy deposition, DSB cross-sections in aqueous DNA remain an order of magnitude lower than in dry DNA. Although our measurements do not resolve molecular-scale dissipation mechanisms, prior theoretical and cluster studies \cite{liu2006collision,markush2016role,domaracka2012ion,barc2014multi,johny2024water,abdoul2015velocity,berthias2015proton,kocisek2016microhydration,richter2018competition,jahnke2010ultrafast,ren2018experimental,wang2020water,milosavljevic2020oxygen,mcallister2019solvation,du2024ultrafast} support the plausibility of energy transfer from DNA to water, mediated by hydrogen bonding and collective modes. Our results indicate that, in condensed phase conditions, the protective role of water dominates over any potential catalytic contribution to damage. This observed directionnal asymmetry should motivate further {\it ab initio} studies to clarify its microscopic origin.

{\it Conclusion.--}Our findings point to a fundamental mechanism by which condensed aqueous environments act as dissipative reservoirs, mitigating direct radiation-induced damage to large biomolecules. By providing {\it in situ} quantification of direct DNA damage across proton penetration depths under controlled chemical conditions, this work establishes a robust experimental basis for future theoretical and {\it ab initio} simulations. These findings also underscore the importance of incorporating solvent effects into MonteCarlo toolkits to improve radiobiological damage modeling in contexts relevant to proton therapy and radiation protection \cite{chatzipapas2023simulation,sakata2020fully}.

It should be noted that our experiments probe DNA under a controlled hydration environment, which differs from the full complexity of the nuclear medium. The nucleoplasmic environment includes additional components, such as proteins, chromatin structure, and ionic conditions, which may further influence radiation-induced processes. Nevertheless, our results provide quantitative benchmarks on the fundamental role of hydration in modulating energy transfer from protons to DNA. 

The methodology developed here offers a versatile platform to study DNA fragmentation dynamics under a range of environmental conditions, including temperature, ionic strength, or radical scavenger concentration. Extending this approach to DNA–protein complexes represents a promising direction for future investigations, potentially providing deeper insights into radiation effects in physiologically relevant contexts.

\begin{acknowledgments}

This project has received financial support from the CNRS through the MITI interdisciplinary programs.

The AIFIRA facility is financially supported by the CNRS, the University of Bordeaux and the Région Nouvelle Aquitaine. We thank the technical staff
members of the AIFIRA facility P. Alfaurt, J. Jouve and S. Sorieul.

\end{acknowledgments}

\bibliography{apssamp}

\end{document}